\documentclass[journal]{IEEEtran}
\IEEEoverridecommandlockouts
\usepackage{cite}
\usepackage{amsmath,amssymb,amsfonts}
\usepackage{algorithmic}
\usepackage{graphicx}
\usepackage{textcomp}
\usepackage{xcolor}
\usepackage{multirow}
\usepackage{makecell}

\usepackage[justification=centering]{caption}
\def\BibTeX{{\rm B\kern-.05em{\sc i\kern-.025em b}\kern-.08em
    T\kern-.1667em\lower.7ex\hbox{E}\kern-.125emX}}
\begin{document}

\title{Data-Driven Machine Learning Techniques for Self-healing in Cellular Wireless Networks: Challenges and Solutions}
\author{Tao Zhang, Kun Zhu, and Ekram Hossain \thanks{T. Zhang and K. Zhu are with College of Computer Science and Technology at Nanjing University of Aeronautics and Astronautics, E. Hossain is with Department of Electrical and Computer Engineering, University of Manitoba, (emails: \{tao,zhukun\}@nuaa.edu.cn, Ekram.Hossain@umanitoba.ca). The work was supported in part by a Discovery Grant from the Natural Sciences and Engineering Research Council of Canada (NSERC).}}
\maketitle

\begin{abstract}
For enabling automatic deployment and management of cellular networks, the concept of self-organizing network (SON) was introduced. SON capabilities can enhance network performance, improve service quality, and reduce operational and capital expenditure (OPEX/CAPEX). As an important component in SON, self-healing is defined as a network paradigm where the faults of target networks are mitigated or recovered by automatically triggering a series of actions such as detection, diagnosis and compensation. Data-driven machine learning has been recognized as a powerful tool to bring intelligence into network and to realize self-healing. However, there are major challenges for practical applications of machine learning techniques for self-healing. In this article, we first classify these challenges into five categories: 1) data imbalance, 2) data insufficiency, 3) cost insensitivity, 4) non-real-time response, and 5) multi-source data fusion. Then we provide potential technical solutions to address these challenges. Furthermore, a case study of cost-sensitive fault detection with imbalanced data is provided to illustrate the feasibility and effectiveness of the suggested solutions.

\end{abstract}


\section*{Introduction}

With the development of cellular networks towards 5G and beyond, they are evolving to more complex structures featured by heterogeneity and dense deployment. In these networks, traditional (e.g. non-automated) methods for network deployment, configuration, optimization, and maintenance will be inefficient and will incur huge operational and maintaining expenditures. This has led to the concept of self-organizing network (SON) advocated by the Third Generation Partnership Project (3GPP) and Next Generation Mobile Networks (NGMN) alliance. SON includes three main functions: self-configuration, self-optimization,
and self-healing. SON capability will enable more flexible planning and deployment of mobile networks, more efficient optimization and maintenance, less manual intervention, lower capital expenditure (CAPEX) and operational expenditure (OPEX) \cite{b1}.

As an important SON functionality, self-healing will automatically detect faults of target networks (e.g. cellular networks) and trigger corresponding actions to fix them. The self-healing functionality mainly includes four phases: fault detection, diagnosis, compensation, and recovery. The goal of fault detection is to find problems such as unacceptable service quality (e.g. due to coverage hole, excessive interference, excessive antenna uptilt or downtilt). Fault diagnosis identifies the root cause based on Key Performance Indicators (KPIs) and alarms. After the faults are identified, recovery actions are launched. To insure quality of service in the process of fault recovery, compensation mechanism is triggered to mitigate degraded network performance in the affected zone through tuning some involved network parameters automatically.

In traditional networks, it is common that operators are aware of service failures only after receiving a large number of user complaints. And for failure recovery, the experience of technicians is of paramount importance. In comparison, the objective of self-healing is to perform these tasks automatically in an active manner. Naturally the introduction of intelligence into networks is required, for which machine learning has been recognized as a powerful tool. Specifically, machine learning techniques are able to automatically generate inference and classification models by training collected data, offering accurate results for reliable decision making. And different types of machine learning techniques (e.g. supervised learning, unsupervised learning, and reinforcement leaning) have been leveraged for self-healing. For example,  many learning algorithms are devised to detect cell outage and to compensate degraded network performance of problematic cell. And some algorithms can train classifiers for fault diagnosis which can discriminate different faults.

However, though machine learning technologies facilitate the development of self-healing methods for cellular networks, several major challenges  exist which can impact the performance and practical implementations. In this article, we classify these challenges into the following five categories:
\begin{itemize}
\item \textbf{Data imbalance:} In cellular networks, due to the occurrence of rare events (e.g. network failure), the collected data sets are usually imbalanced. These imbalanced data can significantly impact the performance of classifiers, which is likely to have a skew towards majority class. However, existing schemes rarely take the issue of data imbalance into account.
\item \textbf{Data Insufficiency:} The insufficiency of high quality data can result in severe over fitting of learning models (e.g., classifier). Firstly, the data set obtained from high-fidelity network simulators may not fully represent the measurements in practical cellular networks. While the real data from network operators (e.g. log data) may not be well organized and labeled. And it is difficult to extract effective information and build knowledge from these data.
\item \textbf{Cost insensitivity:} Misclassification is unavoidable. However, most existing schemes pursue a low detection error rate, while ignoring the fact that different type of misclassification errors can cause different losses to the operators. In such case, considering accuracy as the only evaluation criterion is defective and cost sensitivity should be considered.
\item \textbf{Non-real time response:} Most existing self-healing schemes do not meet the real-time response requirements due to their reactive characteristics. Specifically, they are mainly based on post operations (e.g. diagnosing after malfunctions occur). How to design proactive schemes to reduce the delay and enable real-time response is challenging.
\item\textbf{Multi-source data fusion:} Theoretically, data from varying levels such as subscriber level, cell level, and core network level can be jointly exploited for achieving better performance \cite{b2}. However, the multi-source data bring difficulties to model construction. Therefore, how to perform multi-source data fusion for self-healing is a challenging issue.
\end{itemize}

The main contributions of this article are three folds. Firstly, we discuss the challenges in data-driven machine learning algorithms for self-healing. Secondly, we provide some potential solution directions for addressing these issues. Thirdly, we provide a case study of cost-sensitive fault detection with imbalanced data to illustrate the feasibility and effectiveness of the suggested solutions.

\section*{Challenges in Data-Driven Machine Learning}

\subsection*{Data Imbalance}

Data imbalance often occurs in machine learning and data mining, when at least one class contains more samples comparing to other classes. For convenience, we term the class containing a large number of samples as majority class, and the class including a relatively small number of samples as minority class. The ratio of the number of samples between minority and majority classes is used to measure the degree of data imbalance. In general, in case this ratio is close to 1, the data imbalance can be negligible. Whereas when the ratio is significantly less than 1, the imbalance may hamper the performance of classifiers significantly.

In self-healing, fault detection and diagnosis can be considered as typical classification problems. Accordingly, existing machine learning based classification methods can be applied for which measurements data during networking operation period are collected to train corresponding classifiers. Note that a cellular network is functioning well during most of the running time, and service failure or degradation appear with a relative low probability. In this case, the amount of collected normal status data overwhelms that of abnormal data, which then generates imbalanced training data.

However, traditional classification algorithms are designed with the premise of balanced data set. When they are applied to fault diagnosis in cellular network which has imbalanced data, the results usually lead to a bias towards majority class, and the classification accuracy for minority class is not satisfying \cite{b3}. Accordingly, data imbalance poses a challenging issue for self-healing which is not considered in most existing work.

\subsection*{Data Insufficiency}

Currently, the lack of high quality data poses great challenges for the application of machine learning based self-healing mechanisms, which could even hinder the development in this area. Specifically, machine learning algorithms usually require adequate training data to train a stable model. Nevertheless, when the training samples are insufficient, the learners are likely to achieve knowledge from peculiar features rather than common features in data sets, which may result in severe over fitting.

Data insufficiency arises mainly due to the following reasons. First, for most researchers in universities and research institutions, acquiring sufficient data from network operators is not an easy task due to privacy and business issues. This can be shown in existing literature in self-healing most of which use data from some high-fidelity network simulators (e.g., NS3, Vienna-LTE, and  LTE-Sim). Though these simulators provide good simulation environment, the data collected via simulations cannot fully represent real network scenarios. Also, mobile network measurement data may be collected by means of third-party signal sniffers or some applications in mobile devices, some measurement data (e.g. fault indicating data) are difficult to collect. Second, network operators have huge amount of operation data which are stored in system logs. However, these data may not be well organized and labeled. Accordingly, extracting effective information is difficult. And the data insufficiency also arises due to limited labels. Compared to labeled data, unlabeled data are characterized by large amount. Annotating these unlabeled data usually require experienced engineers and is time/cost consuming and in some cases it may not be always feasible. Therefore, applications of machine learning algorithms for self-healing need to address the challenge of training a model from insufficient real-world data.

\subsection*{Cost Insensitivity}

In order to evaluate the performance of a machine learning method, metrics such as accuracy, generalization ability, interpretability, time and space complexity, as well as cost-sensitivity need to be taken into account. However, the traditional machine learning methods for self-healing focus primarily on achieving maximized accuracy and ignore the cost involved in the classification process (i.e. assume equal costs for different misclassification errors). However, in real-world scenarios, different misclassification errors often have varying costs. For example, within the process of cell failure diagnosis in self-healing, the cost of mistakenly diagnosing a malfunction to a fault-free case is large than that of identifying a fault-free case to be a case for malfunctioning. Detecting a fault-free case as a malfunction at least can attract the attentions of engineers and makes them take actions to check the failure. However, it means that the network fault is neglected when diagnosing a fault as a normal case mistakenly. Thus, it is unreasonable that different misclassification errors are assigned to equal costs in self-healing.

\subsection*{Non-Real-Time Response}

Existing self-healing mechanisms cannot meet the real-time response requirements for future mobile networks due to their reactive characteristics. This is due to the fact that they depend on post operations (e.g. detecting and diagnosing after malfunctioning occurs) which can lead to poor service quality for subscribers. For real-time response, the network needs to be fully aware of the changes of context, so that timely response can be taken when network degradation or malfunction occurs.

\subsection*{Fusion of Multi-source Data}

Currently, for self-healing, cell-level data are frequently utilized for detecting, diagnosing, and recovering from network faults, as well as performing compensation during performance degradation period. Theoretically, data from different levels of sources such as subscriber level, cell level, and core network level can be jointly exploited to achieve better performance \cite{b2}. For example, subscriber-level data (e.g. connection and drop rate, throughput, and delay) are collected from diverse user devices and reflect the communication quality at the user side as well as users' communication behavior/pattern. Operator-level data (e.g. Minimization of Drive Test (MDT) reports, received interference power, and Channel Quality Indicator (CQI)) are collected by the Operation and Maintenance Center (OMC) to monitor the changes of network.

However, the data from multiple sources are characterized by different modality and granularity. Also, there could be ambiguity and spuriousness. Accordingly, these multi-sourced data cannot be directly exploited by most existing machine learning algorithms for self-healing. And how to process multi-source data and adjust corresponding algorithms to achieve the potential benefits of data fusion is a challenging issue.

\section*{Solution Approaches}

In this section, we will present several potential approaches for addressing the above challenges in application of machine learning techniques for self-healing. A concise description of the solutions is given in TABLE I.

\subsection*{Data-Imbalance Solutions}
We will present the following two types of solution approaches for handling data-imbalance problem: data preprocessing and algorithm modification.

\begin{table*}[htbp]
\caption{The solutions for different challenges on machine learning in self-healing}
\begin{center}
\setlength{\tabcolsep}{0.5mm}{
\begin{tabular}{|c|c|c|c|}
\hline
        Challenges          &         Solutions          & Methods & Brief Description \\ \hline
\multirow{7}{*}{Data imbalance} & \multirow{4}{*}{Data preprocessing} & Oversampling & Duplicating minority class data\\ \cline{3-4}
                  &                   & Undersampling &  Removing partial majority class data  \\ \cline{3-4}
                  &                   & Oversampling+undersampling &  Alleviating the problems caused by using one of them \\ \cline{3-4}
                  &                   & SMOTE and its variants & Combining resampling and other methods e.g., KNN \\ \cline{2-4}
                  & \multirow{2}{*}{Algorithm modification} & One-class & Estimating a boundary in each class \\ \cline{3-4}
                  &                   & Hybird strategy & Combining learning algorithms with data preprocessing\\ \cline{3-4}\hline
\multirow{6}{*}{Data insufficiency} & \multirow{2}{*}{Data preprocessing} & Oversampling & Duplicating the samples in each class \\ \cline{3-4}
                  &                   & SMOTE and its variants & Combining resampling and other methods e.g., KNN \\ \cline{2-4}
                  & \multirow{3}{*}{ Using unlabeled data}  & Active learning & Selecting useful unlabeled data and annotating them artifically\\ \cline{3-4}
                  &                   & Unsupervised learning & Dealing with the clustering problems of unlabeled data\\ \cline{3-4}
                  &                   & Semisupervised learning & Combining labeled and unlabeled data\\ \cline{2-4}
                  & \multirow{1}{*}{Algorithm modification} &  Transfer learning& Transferring learning tasks between target and source domains\\ \cline{3-4}
                  \hline
\multirow{3}{*}{Cost insensitivity} & {Cost sensitive learning} & Misclassification cost sensitive learning  & Setting costs for different classes or samples \\ \cline{2-4}
                  & \makecell[tc]{Introducing new \\ evaluation metrics}& \makecell[tc]{F-measure, G-mean, ROC \\ precision-recall and cost curve} & \makecell[tc]{Combining different components of confusion matrix \\ to estimate system performance comprehensively} \\ \hline
{Non-real time response} & Proactive response & Proactive context self-healing & Upgrading existing self-healing from reactive to proactive response \\ \cline{3-4}\hline

\multirow{5}{*} {Multi-source data}& \multirow{5}{*}{Data fusion} & \makecell[tc]{Probability based \\ methods} & \makecell[tc]{Obtaining consistent information \\ from random variables, events, or process} \\ \cline{3-4}
                  &                   &  \makecell[tc]{The theory of evidence \\ based methods} & \makecell[tc]{Using symbolic variables and combination rules \\ to infer consistent information from multi-source data}\\ \cline{3-4}
                  &                   & Artificial intelligence based approaches & Offering a good solution for large-scale complex data\\ \cline{2-4} \hline
  \end{tabular}}
\end{center}
\end{table*}

\subsubsection{\textbf{Data preprocessing}}
It aims to convert imbalanced data to balanced ones through changing the distribution of target data sets before they are fed to the machine learning algorithms. The common preprocessing methods are under-sampling and over-sampling which change the distribution of training samples. Specifically, under-sampling is used to remove several majority class samples randomly and  over-sampling is used to duplicate minority class samples, till a balanced data set is produced. However, under-sampling may result in some important information in majority classes to be lost, and over-sampling may result in over-fitting due to the duplicating operations of minority class samples \cite{b3}. One method to mitigate this problem is to combine under-sampling with over-sampling to achieve a trade-off between less information loss in under-sampling and less severe over-fitting in over-sampling. Another option is to enhance the diversity of samples through combining resampling and other techniques such as $K$ Nearest Neighbor (KNN). One common method is synthetic minority over-sampling technique (SMOTE), which produces more samples for minority classes through computing and inserting new instances among randomly selected data and their $K$ nearest neighbors \cite{b3}.

\subsubsection{\textbf{Algorithm modification}}

The classification problems for imbalanced data set also can be addressed through improving existing machine learning algorithms. Two kinds of solutions may be effective. One is to set reasonable classification boundaries for minority class samples. One-class classifier is a common method for this solution, which estimates a boundary to encompass as sufficient data as possible in each class while minimizing classification errors caused by outliers \cite{b4}. The other is to combine learning algorithms with other technologies such as resampling and cost-sensitive learning. For instance, the combination of traditional classifiers (e.g., Support Vector Machine (SVM), decision tree) and resampling technologies is a good method to improve the classification performance for imbalanced data \cite{b3}.

\subsection*{Data-Insufficiency Solutions}

These solutions fall into three categories: data preprocessing, algorithm modification, learning with sufficient unlabeled data.

\subsubsection{\textbf{Data preprocessing}}
The issue of data insufficiency can still be addressed through generating more data artificially. In this context, some methods used to tackle the problems of data imbalance such as random over-sampling, SMOTE and its variants are suitable to cope with the data-insufficiency problem.

\subsubsection{\textbf{Algorithm modification}}
In the algorithm level, the common solutions are to combine data preprocessing with existing machine learning algorithms. Besides, the concept of  {\em transfer learning}, which is based on the idea of acquiring knowledge from one problem/field (source domain) and adopting them to the learning tasks for a new problem/area (target domain), can be a promising solution approach to overcome the problem of data insufficiency \cite{b5}. In self-healing, some learning tasks are similar to the ones in other networks. For example, there could be enough data available from industrial or wireless sensor networks related to the tasks such as error recovery and intrusion detection. These data could be used to train a learning model and transfer it to a machine learning model for self-healing.

\subsubsection{\textbf{Learning with sufficient unlabeled data}}
When there are limited unlabeled samples and a large number of unlabeled data, the three following solution approaches can be used: active learning, unsupervised and semisupervised learning. Specifically, unsupervised learning deals with the clustering problems of unlabeled data. Semisupervised learning is used to promote the stability of learning models by combining labeled and unlabeled data. Active learning employs learning and selecting engines to find the most useful unlabeled samples, which can be manually annotated to achieve more labeled data \cite{b6}.

\subsection*{Solutions to the Cost-Insensitivity Problem}
These solutions assign distinct costs to different classes or each sample within the processes of learning and decision making, so that the classifiers pay more attentions to costly classification results. These solutions can be classified into two types: cost-sensitive learning and introducing new evaluation metrics.

\subsubsection{\textbf{Cost-sensitive learning}}
In general, the classification costs are assigned either for different categories or for each sample, which are known as class-dependent and example-dependent cost, respectively \cite{b7}. The class-dependent cost denotes that different classes have distinct costs while each sample in one class has equal cost, whereas example-dependent cost implies that each sample has a different cost even when these samples belong to the same class.
\begin{itemize}
\item \textbf{Class-dependent cost:} The methods based on class-dependent costs \cite{b7} allocate different penalties for different classes. These algorithms are mainly classified into two groups. One is to embed this cost into common classifiers like SVM, to achieve more cost-sensitive classifiers. Another is based on Bayesian decision theory to achieve minimum misclassification costs through minimizing conditional risk.
\item \textbf{Example-dependent cost:} The approaches based on example-dependent costs aim at changing cost-sensitivity learning tasks into cost-insensitivity ones by two types of methods. One is altering the distribution of original samples based on different weight values. The other is to optimize the weight values of weighted original samples to obtain expected minimum classification costs \cite{b7}.
\end{itemize}

\subsubsection{\textbf{Introducing new evaluation metrics}}
In the presence of data imbalance or cost insensitivity, evaluating an algorithm using accuracy only is not enough. Instead, more effective metrics should be adopted and such measures include F-measure, receiver operating characteristic (ROC) curve, precision-recall curve, and cost curve \cite{b3}. These evaluation criteria use true positive, false positive, true negative and false negative to reflect system performance from different angles. For example,the ROC curves take true and false positive rate into account through placing them in the same coordinate system.

\subsection*{Solutions for the non-real-time response}

The solution for the non-real-time response problem is to upgrade the existing self-healing approach from the reactive to the proactive one. To this aim, in Fig.~1, we introduce a proactive context-aware self-healing framework. The core idea of this framework is that, the system predicts changes of near-term network performance by using real-time models to capture knowledge from historical and current context information. When possible  faults in near-future are predicted, it can trigger self-healing to adjust the current network parameters so that possible loss caused by the faults is minimized.

\begin{itemize}
	\item \textbf{Data collection:} This block is primarily used to gather enough context information, which are classified into three groups: network context, user context, and device context \cite{b8}.
	
	\item \textbf{Data preprocessing:} The raw data gathered from different context are not directly used by prediction models since they have several characteristics such as redundancy and different granularity. The goal of this step is to change these imperfect raw data into available ones by using methods such as filtering, ranking and fusion.
	
	\item \textbf{Context predictive model:} The main task of this block is to build prediction models (e.g. regression models) with the processed data.
	
	\item \textbf{Self-healing:} This block is mainly used to analyze the predicted results from the prediction model and generate corresponding actions for recovery or compensation.
	
	\item \textbf{Dynamic response:} This block is used to perform actions for self-healing by generating new parameters to be fed into the network to reconfigure it.
\end{itemize}

\begin{figure}
	
	\includegraphics[width=3 in]{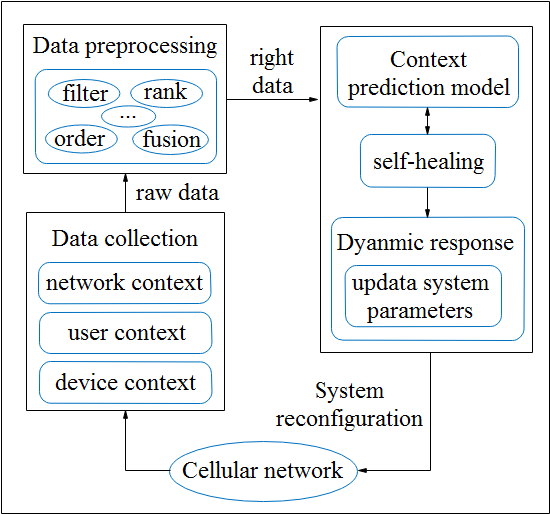}
    \centering
	\caption{A proactive context-aware self-healing framework}
\end{figure}

\subsection*{Solution to the fusion problem for multi-source data}
Fusion of multi-source data aims at obtaining unified information by analyzing and reorganizing the data which come from heterogeneous devices and different scenarios. According to \cite{b9}, we classify the solutions for the data fusion problem into three types: probability based methods, the theory of evidence based methods, and artificial intelligence based approaches. Briefly, probabilistic techniques (e.g. Bayesian analysis and Markov Chain) are frequently utilized to discover consistent information from random variables, events or process, which makes them suitable for dealing with uncertain and imprecise multi-source data. The theory of evidence based methods usually use symbolic variables and combination rules to infer consistent information from multi-source uncertain data. Moreover, the artificial intelligence approaches (e.g. machine learning, fuzzy logic, and genetic algorithms) are used for data fusion due to their strong ability on processing large-scale complex data.

\section*{Case Study: Cost-Sensitive Fault Detection with Imbalanced Data}

We provide a case study on fault detection in order to illustrate the challenges and solutions related to data imbalance and cost-sensitivity issues in machine learning-based solutions for self-healing. We propose a mechanism which enables fault detection through discriminating fault and fault-free measurements jointly considering data imbalance and cost-sensitivity.

\subsection*{Classification from Imbalanced Data}
To handle the problem of data imbalance in fault detection, re-sampling techniques can be used to preprocess the data. However, as mentioned before, the under-sampling may result in some important information in majority classes to be lost due to its removing operations and the over-sampling may result in over-fitting due to the duplicating operations of minority class samples. SMOTE as an improved scheme can generate new minority class samples by means of neighboring samples. And the neighbor samples are selected by the $K$-nearest neighbor algorithm. Therefore, SMOTE can avoid the over-fitting problem. With this method, a new minority sample could be obtained as follows: for a sample $X$$_i$ in minority class, find the K-nearest neighbors to $X$$_i$, then randomly select a sample $X$$_j$ from above neighbors, calculate their difference $diff=X_i-X_j$, and finally obtain a new sample by $X_n=X_i+rand(0,1)*(X_i-X_j)$. In this experiment, we employ following two methods to demonstrate the necessity to consider data imbalance for machine learning algorithms in self-healing and experimental results are shown in Fig.~3. Also, the whole process of fault detection is shown in Fig.~2.

\begin{figure}
	
	\includegraphics[width=3 in]{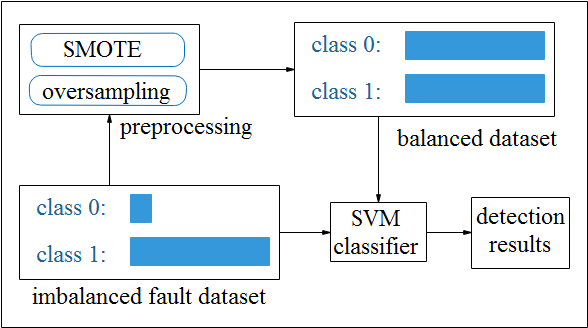}
    \centering
	\caption{Fault detection for imbalanced data}
\end{figure}

\begin{itemize}
\item \textbf{Method 1:} We use SVM to classify the imbalanced data set directly.
\item \textbf{Method 2:} First, oversampling and SMOTE are used to preprocess the imbalanced data set to convert the imbalanced data to balanced data. Next, SVM is used to classify the balanced data set.
\end{itemize}

\subsection*{An Example of Cost-Sensitive Learning in Self-healing}
We will explain the necessity of considering cost-sensitivity for the existing machine learning algorithms in self-healing and show changes of classification results under different costs. We use $C_{ij}$ ($i,j \in \{ 0,1\}$) as the cost of misclassifying true class $i$ to predicted class $j$, where we preset $C_{00}=C_{11}=0$ and $C_{10}=1$. Class 0 and class 1 represent fault and fault-free classes, respectively. The cost ratio denotes the ratio of $C_{01}$ and $C_{10}$. We have done two tests, which are described as follows.

\begin{itemize}

\item \textbf{Test 1:} At first, we use SVM to train a model with training set, and utilize cost-sensitive SVM (CS-SVM) \cite{b10} to train different models based on varying cost ratios (i.e. changing $C_{01}$ from 1 to 30). Also, we combine CS-SVM with SMOTE. Next, the test set is utilized to validate the model, and in this process, misclassification costs are calculated through comparing the predicting labels with test set labels. Finally, the total costs along with different cost ratios are achieved and experiment results are shown in Fig.~4.

\item \textbf{Test 2:} We compare the classification results of CS-SVM under setting different cost ratios, which are shown in Fig.~5.

\end{itemize}

\subsection*{Results}
We use the simulation scenario proposed in \cite{b11}. We only consider binary classification problem in this article. An imbalanced data set is utilized and there are 117 faults data and 3363 fault-free data in class 0 and 1, respectively. We split the entire data into a training set (including 2783 data) and a testing set (including 696 data), and each data is composed of seven key performance indicators (KPIs): retainability, handover success rate, Reference Signal Received Power (RSRP), Reference Signal Received Quality (RSRQ), Signal-to-Interference-plus-Noise Ratio (SINR), throughput and distance. For performance evaluation, we  show the results through ROC curves and use the area under the ROC curve (AUC) to compare different classification algorithms. The larger the AUC is, the better is the classification performance.

For the classification based on balanced fault data, the corresponding results are shown in Fig.~3. As can be seen, compared to method 1, method 2 achieves a higher AUC. This demonstrates that, when the data are imbalanced, the performance of traditional classifiers are tempered, and preprocessing imbalanced data via oversampling and SMOTE is an effective method to improve it. In addition, comparing oversampling with SMOTE, the latter works better. This illustrates that SMOTE can improve the performance of random resampling to some extent.

\begin{figure}
	
	\includegraphics[width=3.2 in]{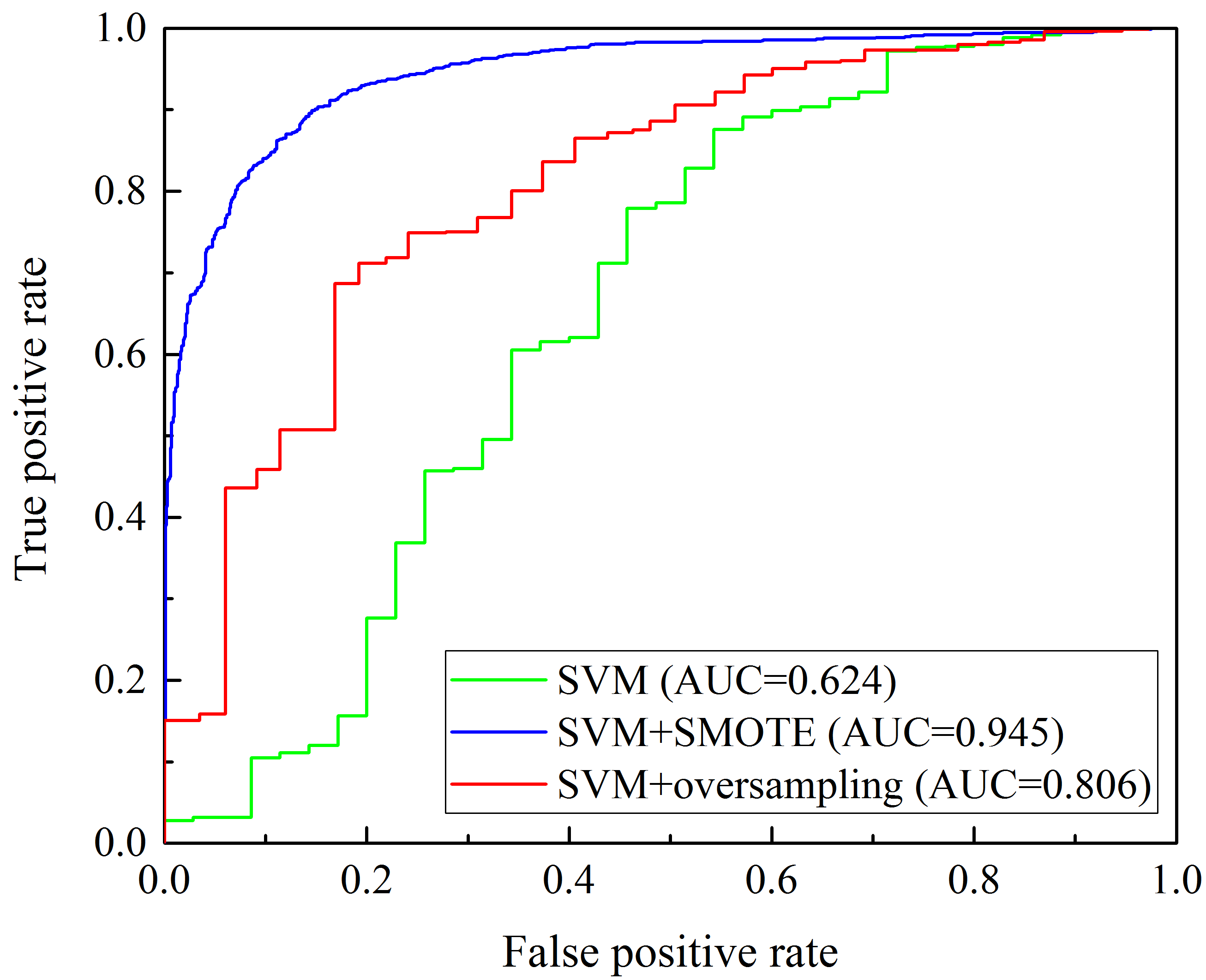}
    \centering
	\caption{Comparison among ROCs for different fault detection methods}
\end{figure}

With regards to the experiment related to cost-sensitivity, as can be seen from Fig.~4, with cost ratio changing from 1 to 30, the total costs of traditional SVM increase linearly while they do not increase for CS-SVM when the cost ratio is larger than 20. Also, lower total costs can be achieved by adding SMOTE on top of CS-SVM. This illustrates that cost-sensitive algorithms can effectively control misclassification results and a hybrid of SMOTE and cost-sensitive learning can provide better results for the classification of imbalanced data. Also, we can see from Fig.~5 that, when presetting a larger cost ratio for CS-SVM, a higher classification performance is obtained. This indicates that the detection of network faults can be easier when setting larger cost ratio.

\begin{figure}
	
	\includegraphics[width=3.2 in]{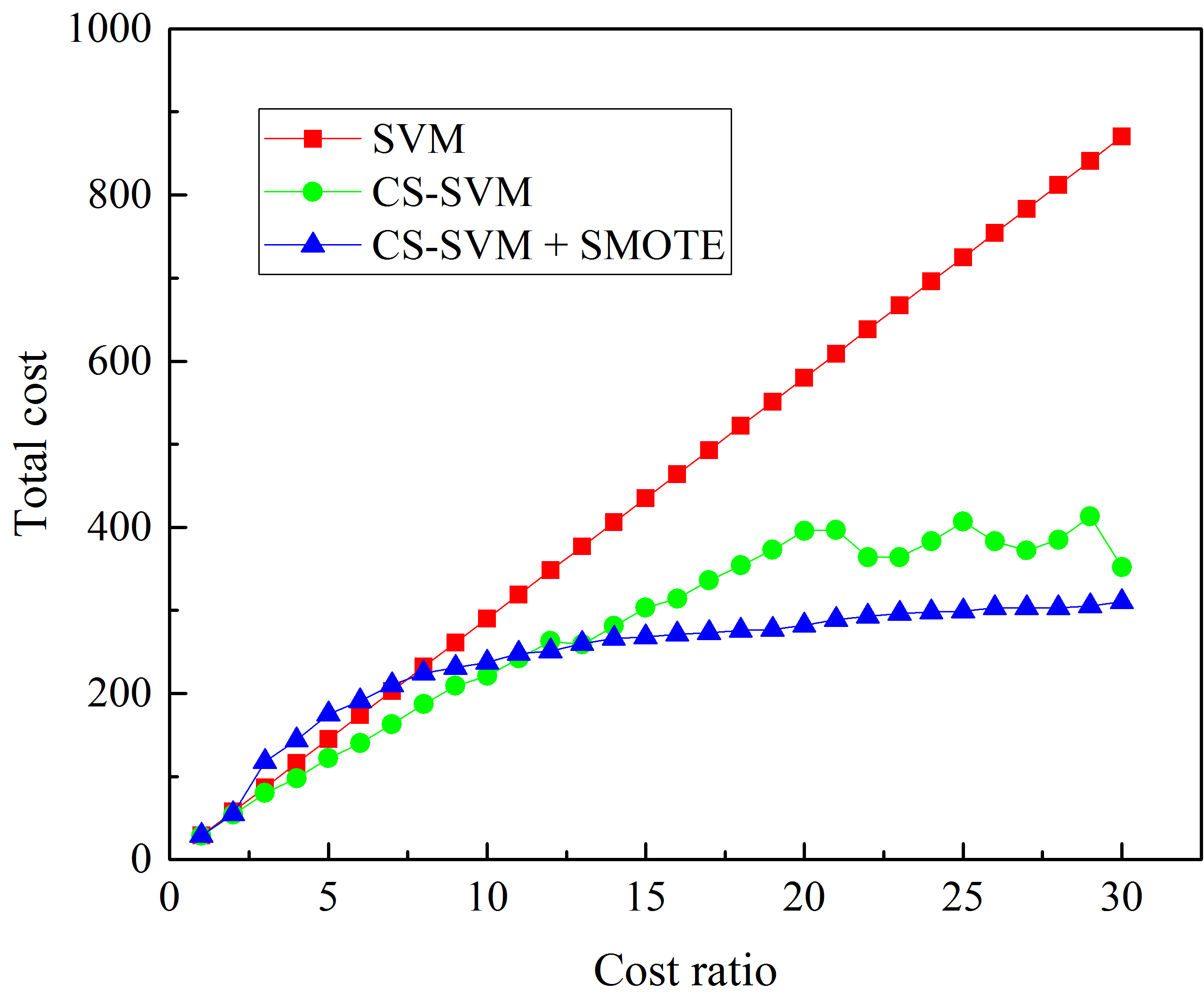}
    \centering
	\caption{Comparison among the methods under different cost ratios}
\end{figure}

\begin{figure}
	
	\includegraphics[width=3.2 in]{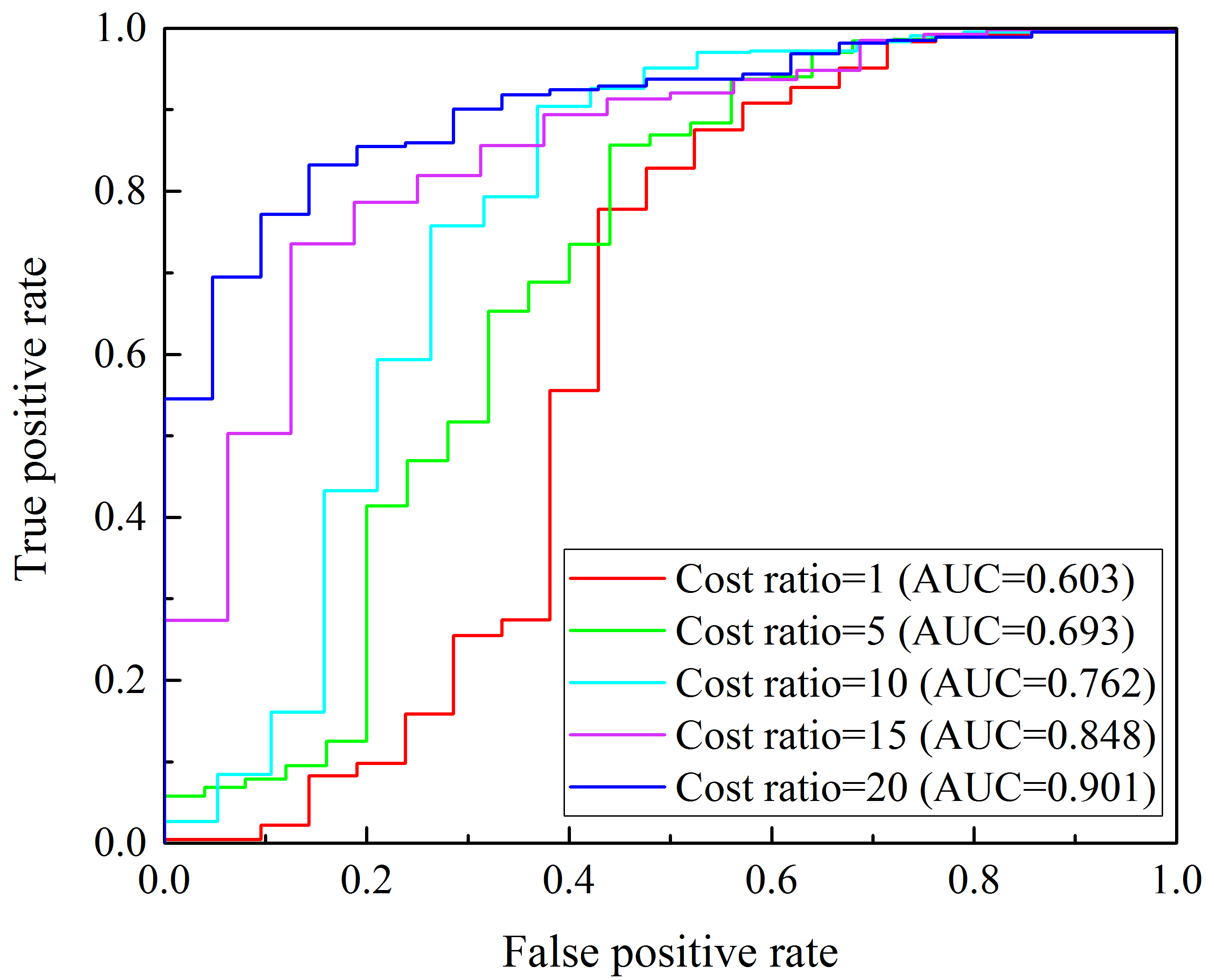}
    \centering
	\caption{Comparison among ROC curves under different cost ratios}
\end{figure}

\section*{conclusion}
Self-healing as a key component in SON will play vital roles in realizing intelligent operation in next generation cellular networks. And it has been well recognized that data-driven machine learning techniques are useful for the development of self-healing mechanism and much research efforts have been put on this topic. However, the application of machine learning techniques in this paradigm faces challenges such as data imbalance and insufficiency, cost-insensitivity, non-real-time response, and the fusion for multi-source data. In this article, we have concisely discussed these challenges and provided potential solutions. Besides, a case study of cost-sensitive fault detection has been presented to illustrate the effectiveness and feasibility of suggested approaches.

\end{document}